 \documentstyle[preprint,aps,tighten,eqsecnum]{revtex}
         \def\be{\nopagebreak[3]\begin{equation}}
         \def\ee{\end{equation}}
         \def\ba{\nopagebreak[3]\begin{eqnarray}}
         \def\ea{\end{eqnarray}}

 \preprint{\vbox{\baselineskip=12pt
 \rightline{ICN-UNAM-02-03}
 \rightline{gr-qc/0203072}}}
 
\begin{document}
 \title{Quantum Geometry as a Relational Construct}
 \author {Alejandro Corichi${}^{1,2}$\thanks{corichi@phy.olemiss.edu},
 Michael P. Ryan${}^{1,3}$ and
 Daniel Sudarsky${}^1$\thanks{sudarsky@nuclecu.unam.mx}}
 \address{1. Instituto de Ciencias Nucleares\\
 Universidad Nacional Aut\'onoma de M\'exico\\
 A. Postal 70-543, M\'exico D.F. 04510, M\'exico\\
 }
 \address{2. Department of Physics and Astronomy\\
 University of Mississippi, University, MS 38677, USA\\
 }
 \address{3. Department of Physics, University of Maryland\\
 College Park, MD 20742, USA\\
 }
 \maketitle
\begin{abstract}
 The problem of constructing a quantum theory of gravity is considered
 from a novel viewpoint. It is argued that any consistent theory
 of gravity should incorporate a relational character between
 the matter constituents of the theory.
 In particular, the traditional
 approach of quantizing a space-time metric is criticized and
 two possible avenues for constructing a
satisfactory theory are put forward.
\end{abstract}
\pacs{PACS numbers:04.60.-m, 04.20.Cv}

 \section{Introduction.}
 \label{sec:1}

  The search for a theory of quantum gravity has been one of the most
 excruciating tasks of modern theoretical physics.  The reasons
 for this  are multiple: To begin with, we have excellent theories to describe
 all other interactions in the standard model of particle
 physics, gravity being the missing item in this approach.
 However, and perhaps even more important is the fact  that the
 realm of quantum gravity seems to confront us with deep conceptual
 problems.  Among many such problems we have diffeomorphism
 invariance
and its connection with observables, the issue of initial conditions for
the wave function
 of the universe, and the problem of time in a quantum theory of
 the space-time, etc.\cite{Conceptual}. Another example is the
 so called information loss paradox associated with the quantum
 evaporation of a black hole.

 The problem of finding a quantum theory of gravity has been
 attacked from several perspectives, the most notable being the
 canonical quantization approach, a la  Wheeler-DeWitt \cite{WDW};
 Loop Quantum Gravity \cite{LoopQG},
 in which there is a natural place for the notion
 of a ``wave function of the universe;" and the String/M Theory
 program\cite{StringT}. The last two have recently enjoyed a certain
 degree of success, in particular by their achievements on one
 front that has been long considered as a ``first test"  that must
 be passed by a candidate theory of Quantum Gravity: The
 identification of the fundamental degrees of freedom and the
 evaluation through statistical mechanical methods  of the entropy
 associated with a Black Hole \cite{bhentropy}.

 However, despite these recent achievements there is a lingering sense
 of frustration in some parts of the gravitational physics community related
 to the fact that not  much progress seems to have emerged from these theories
 in addressing many  of  the outstanding conceptual problems mentioned above.

 We usually feel that there will be a need  for a quantum description of
 gravity only when the  energy scales of the interactions reaches the Planck
 scale. This seems to be a very simplistic view, first because the technology
 available is reaching the point where certain type of experiments once
 thought to
 belong  only in the {\it Gedanken } realm of Einstein, are now  becoming
 possible: Can we test the equivalence principle with an object in a state
 of quantum superposition of different energy eigenstates? Yes
 \cite{Quant Free Fall}. Can we hope to see quantum  fluctuations in the
 space-time metric? Yes \cite{Quant Fluct}. The list of such
 experiments is increasing
 and it seems reasonable to expect that in the near future we might be able
 to perform experiments with quantum mechanical sources of the gravitational
  field (such as a SQUID carrying a superposition of currents in opposite
 directions
 \cite{Ahluwalia}).
 Moreover,  we would like to have, at least in principle,
 a way to envision a a manner to describe space-time at, say, the atomic scale,
 whose curvature is associated with  a nucleus which is in a state  for which
 the position is not well defined. There is certainly no
 Planck scale involved here!
 Note that this last issue could be accessible to experiment in setups
 analogous to those considered in \cite{Penrose}.

 The purpose of this article is to suggest a new approach to this issue
 that centers
 on the conceptual problems  underlying our notions of
 space-time in view of the quantum nature of the matter that inhabits it.
 The hope is that  by doing so we might be guided towards a description of
 gravity at the quantum  level that  will be able from the outset to deal
 with the conceptual problems on which little light seems to be being shed
 by the most popular approaches to the subject.

  In this paper we will take the view that gravity is {\it NOT LIKE} the other
 forces in nature, basically because  gravity describes the stage where
 the rest of the physics take place  while the other forces are just actors
 on this stage.  The previous statement  is meant to appeal to our
 intuition, and should not be taken in the literal sense, among other
 reasons because, as we will later argue on this paper,  the stage is in a
 sense just a set of relations between the actors. A more precise  way to
 argue that gravitation is not just another force is to note that  gravity
 alone  is responsible for determining the causal structure and thus  the
 commutation properties of other fields.

 The motivation of this work is the following question: How would we have
 arrived at a description of gravitation if our  normal physical experiences
 came  from the quantum world, and the classical limit  had not been
 discovered. More  precisely,  what is the analogue of the conceptual
 path that took us from special to general relativity, if the starting point
 was  a quantum rather than a classical theory?  Much has been said  about
 the possibility that both quantum theory and general relativity might need
 modifications when we  want to bring them together into a quantum theory of
  space-time.
 \footnote{ideas in this direction have been explored within the string theory
 perspective in \cite{witten}.}
   The ideas  that will be presented here constitute an example
 of one such approach, one which  falls in the category of  those based on
 relational ideas, several of which  have been considered previously
 \cite{relational} but  as far as the authors know, none of  them
considered the issue in the light we suggest in this work. We
 will elaborate on this view in the rest of the paper. In
 Section~\ref{sec:2} we
 will describe the general considerations that  we take as guiding
 our search for a framework  with which to start the construction of a theory
 of Quantum Gravity. This is the core of this paper and the remainder is
 presented to help to envision the type of  construct we expect to be
 appropriate . In Section~\ref{sec:3} we will sketch a proposal based on the
 generalization of the algebraic approach to quantum field theory. In
 Section \ref{sec:4} we will suggest a line based on the relativization of the
 notion of quantum state as proposed by C. Rovelli \cite{Rovelli1}
 and others.
 We  will end  in Section~\ref{sec:5} with a brief discussion
 centered on the  main objections that could be raised to such a program.

 \section{Geometry and Physics}
 \label{sec:2}

 Here we want to analyze the notion of {\it Physical Geometry}, that is the
 geometry that  we associate to the physical world in a way that
 implicitly shows the contrast with idealized mathematical geometry.

 Let us start by considering what is the role of geometry in classical (as
 opposed to quantum) physics. More precisely we want to analyze the
 meaning of such  assignment of geometry to our physical world. We have
 learned from Einstein that we measure intervals between events. In order
 to do this we must first identify these events, by for example singling
 out one point  on the world line of a point-like object, which can be
  done for example by considering the intersection of two  such world
 lines (We can be thinking of the ``passing of an object  in front of a
 particular observer", or the emission of a light pulse from a certain
source), and second we must use a physical device to actually measure
 the interval.  We would use a ``clock" if the interval is time-like,
 a ruler if the interval is space-like, or define it as null if the two
 events can be connected by a light signal (to be precise we must ensure
 that the measurement is done along the prescribed interval).  In all
 these cases we need physical objects, clocks, rulers, or light pulses,
 in order to be able to define what the geometry is. This is not just
 saying the obvious, that we need such objects to measure the geometry.
 What we want to stress is that we need to specify these physical objects
 in order  to {\it define} what we mean by  ``the geometry of our physical
 world".  We note for example that in principle it is not {\it  a priori}
 certain that the geometry could not depend on the objects we choose to
 employ to define it. For instance we could choose to measure spatial
 intervals with a ceramic rods or with  metallic rods, and imagine a
 situation in which the temperature  trough the region  being examined
 is not uniform.  Under such circumstances the straightforward
 determination of the geometry of the region would give different results.
 For instance, the geometry determined with  the ceramic rods could be
 flat while the one determined with the metallic rods could  have nonzero
 curvature. This is all obvious, and everybody would tell us that the
 use of the ceramic rods is the correct choice in this case because of
 the large thermal  dilation coefficient  of  the metal . Actually we
 would be told that even if we use ceramic rods we need to correct for
 such thermal effects, and only then we would obtain  a determination of
 the ``true geometry" which would be equivalent to that  which is
 determined by means of ideal ``length preserving"  rods. One would normally
 expect to see the quotation marks of the previous phrase on the word
 ideal, however, what  we want to point out is that,  there is in
 principle a problem of  establishing what is a ``length preserving"
 object. How would we know, even in principle when an  object  does
 not change its length?. By measuring its length in  different
 circumstances? Yes!  However, in order to measure its length we need
 to use a length preserving object! We must emphasize that the issue
  is not entirely technical.  We could consider ---and in practice
 we do--- replacing standard physical objects by generalized objects:
 actual physical objects complemented by well defined prescriptions
 of how to correct for the changes in their  assigned lengths under
 specific circumstances (such as taking into account the corresponding
 thermal expansion coefficients and  accompanying every length
 determination with a simultaneous temperature determination).  We would
 still be faced with the issue of  having to define the operational
 procedure to  determine the length preserving  assignment to a physical
  object or a generalized object. We could envision  the consideration
 of  more modern methods, as  for example, choosing to measure an
 object's length with light signals and clocks, while taking the speed
 of light as 1 by definition. This only converts our  problem into that
 of choosing clocks that run at a ``constant" rate.  It is evident that
 this faces us with complete analogous conundrums. How would we measure
 the rate of ticking of a clock to determine whether it is constant
 without the use of another clock?

  In classical physics we are in effect ``solving"  these
 dilemmas by the use of  judicious definitions which  make use of the
 following empirical fact, F1: {\it There exist objects and clocks
 (usually generalized objects and clocks) that when used  to define
 the lengths of intervals, result in empirical laws of physics that
 are particularly simple (as for example, the law of inertia)}
 \cite{Reichenbach}.  In this way we select the objects that give
 an empirical content to our assignment of geometry to the classical
 level of description of our world. What are the objects, actually
 the class of objects $\cal O$, that give a meaning to the geometry at
 the quantum level? Does such class of objects exist at all? Are there
 different classes of objects, say  ${\cal O}_1$  and ${\cal O}_2$,  each
 leading to equally simple laws of nature, which however are different
 for  different  selections of the class of objects? In this case we
 would confront head on the problem of  having two different  geometries
 for the same physical situation, one associated with the class ${\cal O}_1$
 and  the other  with the class ${\cal O}_2$!  We will not attempt to give
  definite answers to  these questions here, but will finish this
 illustrative sequence by a question for which we expect the answer
 to be negative: If a class of objects can be so chosen,  would we
 expect it to contain enough objects to give meaning to geometry at
 all scales?  Note that the limiting scale could be coming from  the
 physics of the objects and not  necessarily from the gravity sector
 alone, that is,  the limiting scale need not be the Planck scale
(See, for instance, the discussion in \cite{AC2} and references therein).

 Recognizing and keeping in mind these points could turn out  to be
 important (and we will argue that they are indeed essential) for
 considerations about the nature of a Quantum Theory of Gravitation.
 In this paper we will take the stand that  the standard views on these
issues rely  on idealizations    that are not unlike the idealizations
 of pre-relativistic physics about the existence of absolute space and
 absolute time, or those in pre-quantum mechanics about the existence of
 well defined  particle trajectories or the existence of a fully
 deterministic physical world. That is , we will take the view that,
 in the same way that those idealizations turned out to be impediments
 for  the advancements towards  Special Relativity and Quantum Mechanics,
 the notion of a physical geometry
 existing independently of the physical objects with which to determine
 it (in the sense of defining it), is an impediment towards the
 construction of a quantum theory of gravitation.

 There are several issues associated with the previous considerations.
 First, we note that, while  at the classical level we must
 consider the objects which are used to define the geometry (i.e
 the objects satisfying F1) as classical objects, at the quantum
 level the natural objects that  should be considered as defining
 the geometry should be themselves ``quantum objects". Thus, given
 that the meaning of  the physical geometry arises  solely from
 statements which must be expressible
  in terms of the objects selected to define it, such meaning should be
 taken to be a certain codification of correlations between physical
 ``events" (this word is used here in an imprecise sense because at
 this point the discussion refers to both the classical and the
 quantum cases) associated with such objects.  At the classical
 level we could be  talking about the number of ticks of a clock
 along the world line segment joining two events with which the
 clock  eventually coincides. At the quantum level we need to
 consider very different sorts of things. One must then expect such
 correlations and their codification to  depend on the type of
 objects one is considering, and therefore the kind of objects
 representing this information (the kind of objects that represent
 geometry) should be expected to be very different in the classical
 and the quantum cases. In particular, we don't expect the quantum
 objects to be described by the same constructs as ordinary quantum
 matter. These differences should be as significant, say, as a
 space-time vector (considered here as describing the state of a
classical particle) is from a wave function (considered here as
 describing a quantum state of a particle).
 Note that up to this point we have not specified what we mean by
``quantum objects''. Our considerations have so far been very general. In
Sections \ref{sec:4} and
\ref{sec:5} descriptions of a more specific character will be given in
the context of the corresponding approaches.
  In considering these issues
we must keep in mind that there are
in  principle two roles played by the physical geometry in a geometric
 theory of gravitation: 1) It is  an entity codifying relationships
 between ``events" defined in terms of the ``matter" content of the
 theory, and 2) it is a dynamical  entity whose behavior is part of
 the subject of the theory.

  Let us try to be a bit more specific at this point by saying not
 what  should be done, but what  should {\it not} be done and why. The
 (most widely accepted)
 classical description of gravitation involves a space-time (i.e. a
 manifold with a  pseudo-Riemannian metric) $(M, g_{ab})$, which
 carries information about the behavior of geodesics, their points
 of intersection,  which would define events, the length of the
 interval between two such events along a  given geodesic, etc. It
 is natural that such an object (i.e. a tensor field) should be used
 to describe the geometry which is defined in terms of point particles
 whose state can be described in terms of a four vector $u^a$ and a
 point $x$ in space-time.
  On the other hand, when the geometry is defined in terms of quantum
 fields [at this point the use of these words should not be taken
 to indicate the standard mathematical  description of such objects,
 but the underlying physical entities  they are thought to
 represent], the physical counterpart of a  geodesic (viewed as
 the world line of a classical test particle) or  the physical
 counterpart of a four vector tangent to a space-time point (viewed
 as a perfectly  localized particle with a well defined four momentum),
  cease to exist! That is, the objects that would give meaning to
 $g_{ab}$ are not part of the realm of physics we are attempting to
 describe. Therefore we should not try to construct a quantum
 counterpart of $g_{ab}$, i.e. a quantized gravitational field
 $\hat g_{ab}$ or any equivalent object, because such a construct
 would be a quantum object carrying information about classical
 correlations! An unholy mixture indeed.
 In fact, doing this, would have
 meant that we  changed aspect 2) of the geometry from classical
 to quantum, but did not change aspect 1)
 \footnote{A remark is in order. The previous discussion suggests that even
 if one already has been successful in
 constructing a quantum theory of gravity, and one  tries to define the
 equivalent of $\hat{g}_{ab}$,
 the task will be  non-trivial or even impossible.}.

 What is needed is to construct an object $\hat{\hat G}$  (the double
 hat is meant to indicate that this is a quantum object but not
 necessarily of the same sort as a quantum field $\hat \phi(x)$)
 which carries information about correlations  of quantum objects.
 The analogy we should use as a guide should be of the following type:
 The metric is an
 object that encompasses information about the way we transport
 (specifically, parallel transport) the description of the state of
 say a classical particle (four vector) from one point to another,
 and thus the quantum geometry should  be described by an object
 that itself  encompasses information  about how to (in some yet
 undefined sense) ``parallel transport" the description of  a quantum
 state of an object from one region to another (or from one
 ``observer'' to another).
  Needless to say, the previous statement should be taken merely
 as a philosophical guiding principle, and any attempt at a specific
  realization should start by stating exactly what the objects are that
 will be used in defining the quantum geometry, what exactly is a
 quantum state of such an object is, and exactly what these
 correlations that the geometry is meant to codify.  Given that an
 ordinary quantum mechanical state of a system has a much richer
 structure than  its classical counterpart (The Hilbert space of a
 one particle system is infinite dimensional while the classical
 phase space is six-dimensional) there is a much larger set of
 possible aspects the correlations describing the geometry could
 refer to. Note also that we are envisioning the existence of  some
 notion of  localization which is in principle absent in the notion
 of a state in quantum field theory, either in flat or curved
 space-time.

 An alternative way of indicating the type of object  that  we expect
 would describe the quantum geometry (i.e. the kind of object that
 $\hat{\hat G}$ should be), is the following:
  We can think of the  metric in general relativity  as a
 codification of how to  join  local inertial frames to cover
 extended regions of space-time. Actually, we know that given the
 metric and a space-time point there is a well defined procedure to
 construct locally inertial coordinates (Riemann normal
 coordinates), and thus by carrying out such constructions at several
  points we can end up with an approximate description of the global
 space-time which would be covered by local inertial  frames,
 whose global properties would be encoded in the rules that allow us
 to translate from one frame to another, and the description of vectors
 in the intersections of the regions covered by two such frames. By
 analogy we  expect  that the quantum geometry  would codify the way
 to translate the description of the states of a quantum systems
 from one ``local frame" to another.

  The next issue we are immediately confronted with is that of the
 notion of physical geometry in the COMPLETE absence of objects.
 That is, we imagine having selected the type of objects that, in
 principle,  give meaning to the physical geometry but would want
 to consider the possibility  of  associating a geometry to a physical
 situation in which these objects are absent . In the classical
 realm what we  mean in these cases is the following: We consider
 the  empirical content of the description of,   say, an empty
 space-time by a certain metric to be associated  to the way that
 ``test"  objects (particles, light pulses, clocks etc) ``would
 behave if placed'' in that  space-time. This presents no particular
 problem in this case because of two very  special features of
 classical physics: First, there is no limit to how closely we can
 approximate, say,  an ideal  test particle  (which would not have
 an effect on the  geometry) by a real physical particle. We can
 make its four momentum as small as we wish without any effect on
 our ability to localize it as much as we want. Second, in
 classical physics the object is unaffected by its  observation.
 It is clear that these two nice features do not survive the
 passage from classical to  quantum  physics: The first feature
 is denied to us by the Uncertainty Principle, and the second
 feature, which  had  allowed us to identify (in the classical realm)
 the observed and unobserved object, must be removed from consideration
 in the quantum realm, because failure to do so often results in
 internal contradictions. That is, we cannot identify, say, a double
 slit experiment in which one of the holes is being monitored for
 the passage of a particle, with a double slit experiment in which
 this monitoring in not being done. We know that at the quantum-mechanical
 level the description of a joint state of two systems
 is not the same as the sum of the descriptions of the two systems.
 Thus we do not expect a viable identification of ``a geometry
 with"  and ``a geometry without" the objects that serve to define it.
 For an earlier discussion of this point see \cite{Ahluwalia:dd}.

 The previous discussion  very strongly suggests that we must abandon
 the notion of assigning a geometry to a physical situation in the
 COMPLETE absence of objects. Moreover,  we believe that the fundamental
 objects describing the matter in our world are not particles, but
 rather fields, and that therefore vacuum is not in any sense ``the
 absence of  everything", but  rather merely a state of the matter in
 the universe, and that, moreover, there is in general   (in particular
 in non-stationary space-times) no state that can be  uniquely
 described by the word ``vacuum".

  Finally  we point out a conclusion that seems to emerge from the
 points we made above: {\it  There should be no place for a
 counterpart of Minkowski space-time, and moreover there
 should be a fundamental limit to the degree to which  such
 space-time can be approached}. Suppose we want to have a region
 with a geometry as close to Minkowski as possible, that is a
 geometry that is as well defined as possible and  as flat as
 possible at any scale.
 In order for it to be as  well defined as possible it
 should be probed with quantum objects in as much detail as
 possible, but in order for it to be as flat as possible we must,
 among other things (here we are guiding ourselves by the
 correspondence principle which in this case indicates that matter
 will be associated with curvature), place as little matter as
 possible in that region, The two requirements are certainly
 incompatible. Thus there should be bounds on the flatness of the
 physical geometry.

 Several remarks are in order: 1) It might seem that the discussion
 above leads to a theory in which Lorentz invariance is violated.
 Certainly this is a possible implication of our arguments. This is
 particularly intriguing, since (local) Lorentz invariance is at the
 heart of our present formulation of physics, needed implicitly in the
 mere definition of spinor fields on space-time. One possibility is
 that Poincare invariance is fundamental at the level of defining
 unitary representations (and thus having the different types of
 possible "particles"), but that this symmetry is not explicitly
 present in the formulation of quantum geometry (just as  Poincare
 invariance is meaningless in general relativity). One would expect
 to recover approximate Poincare invariance at some scale. 2) We
 have not touched upon the validity of the equivalence principle.
 It is our understanding that, as presently formulated, the
 equivalence principle refers to classical objects, and thus
 possible violations can be expected when quantum objects come into
 play. On the other hand, we expect that there should be some local
 validity of the equivalence principle, which in ordinary terms is
 achieved when the quantum system is localized enough as to be in
 an idealized (local) inertial frame. Finally, let us remark that
 it would be desirable to have a quantum replacement for the
 equivalence principle that would serve us as a guiding principle
 in the search for a quantum theory of gravity, just as the
 classical equivalence principle led Einstein to general
 relativity.

 \section{An Approach Based on the Algebraic Formulation of Q.F.T.}
 \label{sec:3}

 The proposal we are going to present here is   very sketchy and
 should only been seen as a possible starting point to develop a
 framework with the characteristics discussed above.
  Its main virtue is that allows us  to present in a relatively
 concrete way the type of construct we have in mind.  The motivation
 of this particular proposal is to remove from the start  one of
 the conceptual problems alluded to at the beginning of this paper:
That of the meaning of the wave function for the universe, for which
 there is by definition nobody that can make the measurements that
 normally give it content \footnote{One view on this issue that has gained
 popularity is associated with
 the decoherence functional etc., but lingering problems exist
 associated with the need to divide the universe into systems playing
 different roles, one that remains the object of the description and
 a second that plays the role of an external medium that induces the
 decoherence. For more detailed criticisms see \cite{fay}.}. The present
 proposal will consider the usual type of quantum  states to have
 in general only local validity and  can, but need not, lead
 to global states. Thus, at least the need for  an object like the
 wave function of the universe disappears from the start. A further
 motivation  for  introducing this feature is the analysis by
 Sorkin \cite{Sorkin} of the applicability of the standard Q.M.
 interpretation of the   measurement problem to quantum field theory,
 which  indicates, that if it is taken literally, leads to
 unacceptable violations of causality (it allows EPR-type experiments
 to actually transmit information).  One basic feature of this
 analysis is the assumption that we can associate a quantum state
 with arbitrarily large
 regions in all circumstances, and thus it seems appropriate to
 call this assumption into question.

  Let us start by reminding the reader that in  the algebraic approach
 to quantum field theory, which  seems to be the appropriate setting
  to consider quantum fields in arbitrary space-times \cite{Wald}, one
 forgoes the notion of a Hilbert space of states, and takes as starting
  point an algebra $\cal A$  (more precisely a $C^*$ algebra) of abstract
 operators (usually thought of as corresponding  to  the smearing
 of polynomials in the field operators by suitable  smooth
 functions) which are identified with the observables of the
 theory, and the states are  then defined to be linear mappings
 $\omega : {\cal A}\to  C$ satisfying a positivity condition
 $\omega (A^*A) \geq 0,  \forall \ A\in {\cal A}$   ( and a
 normalization  condition  $\omega(I)=1$ ), and where the meaning
 of $\omega (A)$ is that of the expectation value of  the
 observable $A$ in the state $\omega$.

  The idea of the approach is to consider localized states in such a way
 as to reproduce locally the setup of quantum field theory, but deny in
 general the existence of global states, and instead introduce rules to
 perform ``parallel transportation" of these states. The complete
 collection of such rules being identified as the  quantum  geometry.

  Let us now present the sketch:
  Let  $M$ be a manifold, and let ${\cal A}$ be a  $C^*$ algebra of
 abstract  operators, each associated with a particular open set $U$
 of $M$.  For every open set $U$ we define the subalgebra define
 ${\cal A}_U$ of local operators associated with $U$   to be the
 collection of all elements of ${\cal A}$ associated with open sets
 contained in $U$. A local state at $U$ is a mapping
 $\omega_U : {\cal A}_U:\to C$ with similar properties as before.

  Two comments are in order: 1) Normally we would have expected
 that once the state at $U$ is  specified it would determine the
 state for all of the domain of dependence $D(U)$ of $U$. However
 stating something like this would presuppose that the causal
 structure is determined. Instead, we would hope to define the domain
 of dependence, and thus the causal structure, in terms of the
 ``correlations" of the states associated with the different regions.
 We would not expect, for example, that the causal structure would be
 always well defined! 2) Note that even though we might have the
 local states defined everywhere in the sense that we might have an
 open  covering  $\lbrace U_\alpha \rbrace, \alpha \in J$ of $M$ and
 an assignment of a local state $\omega_{U_\alpha}$ for every element
 of the covering, we might still not have a well defined global state!
 We might lack information about correlations.

 Next we would need an axiom assuring that the local  laws of physics
 are ``everywhere" the same . We can try something like this: For all
 $p, q \in M$ there exist neighborhoods
 $U_p$, of $p$  and $U_q$ of $q$ and a mapping
 $\phi : {\cal A}_{U_p} \to {\cal A}_{U_q}$ which is an
 algebra isomorphism.  This mapping need not be unique.

 Next we would describe the geometry. The idea is to generalize
 the notion of connection.
  We do not want however to consider parallel transport along a
 line, nor do we have objects associated to points that we might
 wish to transport along such a line. The natural  generalization
 seems to be the assignment of the notion of parallel transport to
 every  one-parameter family  of diffeomorphisms that takes
 an open set into another continuously. A  quantum
 geometry for the manifold $M$
 will be an assignment of a mapping $\Psi_f : {\cal A_U}\to {\cal A_U}$,
  to every one-parameter family of diffeomorphisms $f(\lambda)$ taking
 an open set $U$ to an open set $V$. This assignment would then be
 a realization of $\hat{\hat G}$

  This setup has touched so far only kinematical aspects. We would next need
 to introduce dynamical aspects as well as self consistency
 restrictions.  The fact that we must at some point be able to
 recover  GR, and the appropriate quantum field theory in flat
 space-time in the corresponding limits,  indicate that the
 dynamics of $\hat{\hat G}$,
 must be associated with the  generalized state of the matter
 fields, and the dynamics of the matter fields must be
 compatible with the causal structure that  in the appropriate
 circumstances emerges from  $\hat{\hat G}$.
 Note that some steps in the direction of defining an algebraic QFT
 in a diffeomorphism invariant  context were given in \cite{rainer}.

 \section{Approach based on the program of Relative States.}
 \label{sec:4}

  In a recent article C. Rovelli \cite{Rovelli1}, in considering anew the
 measurement problem in Quantum Mechanics, in particular the collapse of
 the wave function in situations  of the Wigner's Friend type, comes to the
  conclusion that the wave function of a system  should not be considered
 as an attribute of the system in the sense of describing the absolute
 physical state of the system, but rather as an attribute of the relation
 of one system (the described system)  to another (the describing system),
 in the sense of encoding  the information that the second system has about
 the first. In this way the problem of collapse of the wave function  which
 is  instantaneous in a  given reference frame, can in principle be
 disassociated from the collapse of the wave function in other frames.
 The scheme has been considered as dealing with interpretational issues
 and was in no way intended to change the standard predictions of Quantum
 Mechanics. We will nevertheless adopt this idea and consider it  as
 possible starting point for a relational scheme of the sort we have been
 advocating . Again, the following description is merely schematic.

  Let S be the collection of all subsystems in the universe and consider
 assigning to every ordered  pair of them $(A, B)$ a density matrix in an
appropriate Hilbert space, representing the information that $A$ has
 about $B$. When   the  corresponding information is non-existent the
 assignment will be  a multiple of the identity matrix.   In general we
 would expect that such a scheme would leave open the possibility of
 neglecting certain important correlations,  for instance $A$ might
 assign a  density matrix  to $B$  and a density matrix to $C$, but be
 unable to assign more than  the identity to $A$ and $C$ together because of
 lack of information about the correlations. We will not presuppose that
 it is always meaningful to talk about a system preserving its identity
 trough ``time" and thus our systems should be considered in general as
 localized in space and in time. We would of course need to make the
 previous statement meaningful  in a way that is internal to the theory
 (i.e., in terms of considerations about correlations, etc.) as we would want
 space-time to be an emergent property of such a setup. Now we can imagine
  another set of objects,  assigned say to every pair of  pairs of the
 form $((A,B), (A,C))$ the correlations needed  for $A$ to construct a
 density matrix associated to $B$ and $C$ together. Let us call these object
 ``Quantum Correlators".  According to our  discussions it should be
 possible to consider  a  (large enough ) set of Quantum Correlators  as
 describing the structure of space-time. It seems that this is indeed the
 case, for if we are given a ``complete" collection of  wave-functions
 and  quantum correlators we should be able to construct the global
 (space-time) states of all the quantum fields and from these, at least
 in favorable circumstances,  one should be able to deduce the
 Space-time geometry.

  Again, we have touched only on kinematical aspects, while clearly a
 dynamical content that ensures self consistency in the assignment of
 states to systems and subsystems, and that allows us to recover
 General Relativity and Quantum Mechanics in the appropriate limits
 needs to be considered.

 \section{Discussion.}
 \label{sec:5}

 In this note, we have argued that  the physical  significance of
 the  geometry of
 space-time can not be separated from the description of the matter
 fields that probe such geometry, and that it is therefore natural that
 a quantum theory of gravity should reflect  this fact. We are well
 aware of possible dangers of adopting a purely operational approach to the
 construction of physical theories, where one might encounter technical
 difficulties and make little concrete progress. Furthermore,  our position
 should not be seen as the same position adopted, for instance,
 by Rovelli and Smolin who have argued
 that loop quantum gravity (or its close relative, the Spin-Foam
 Models \cite{spinfoam})
 are {\it already} examples of relational quantum theories of gravity
 \cite{relational,rovelli2}, {\it in the vacuum sector}.

 We should also point out that we have left an important issue
 almost untouched. Namely, one expects that there is a scale in
 which the quantum description of the geometry becomes necessary.
 So far, we have not made a concrete proposal about where this
 scale can be found. In the standard approaches, one assumes that
 this scale will be the Planck scale (in the canonical approach) or
 the stringy scale (lately, theories formulated with ``large extra
 dimensions'' put this scales at lower energies). From the
 dimensional point of view, these seem to be the most natural
 choices. A new proposal for a relevant scale should, in our view,
 arise from the dynamics of the quantum objects involved in the
 formulation (See \cite{AC2} and references therein).

 Our point here is that the development of  the physics
 of the last century  presented us with several instances in which our
 attachment to certain  idealizations based in  our previous physical
 experience turned out to be an impediment to progress.  We have made the
 case that the notion of a geometry that  exists independently of the
 objects used to  define it  might
 be one such impediment.  We have also pointed out  that there are several
 outstanding problems associated with the standard interpretation of
 quantum theories that are particularly exacerbated when considered
 in a general relativistic setting, and have thus tried  to  formulate
 approaches  that give some hope of helping in the resolution of  those
 issues. After all, if  quantum gravity is finally conquered and it does
 not help us with these interpretational problems of Quantum Mechanics,
 then it is very difficult to envision from whence clues  for their
 resolution might come.
 We have thus advocated a relational point of view and have suggested a
 few directions that could be tried for the implementation of
 these ideas.

\section*{Acknowledgments}

 AC would like to thank Luca Bombelli for a careful reading of the
 manuscript. We would like to thank the referees for helpful comments.
 We would like to acknowledge partial support from
 DGAPA--UNAM Project No. IN121298 and CONACYT grants 32272-E and
 J32754-E. This work was also partially supported by NSF grant  No.
 PHY-0010061.

 \end{document}